\documentclass{elsart}                            \usepackage{latexsym}
\usepackage[final]{epsfig}  

\newcommand{\be}{\begin{equation}}     \newcommand{\ee}{\end{equation}}
\newcommand{\beas}{\begin{eqnarray*}}
\newcommand{\eeas}{\end{eqnarray*}}
\newcommand{\bea}{\begin{eqnarray}}   \newcommand{\eea}{\end{eqnarray}}
\newcommand{\req}[1]{(\ref{#1})}
\begin{document}
\begin{frontmatter}
\title{Matching games with partial information}
\author[]{Paolo   Laureti\thanksref{ePL}},  \author[]{Yi-Cheng
Zhang}
\address{  Institut  de  Physique Th\'eorique,  Universit\'e  de
Fribourg, P\'erolles, CH-1700 Fribourg, Switzerland }
\thanks[ePL]{paolo.laureti@unifr.ch}
\date{\today}
\begin{abstract}
We analyze  different ways of  pairing agents in a  bipartite matching
problem, with regard to its scaling properties and to the distribution
of individual  ``satisfactions''. Then we  explore the role  of partial
information  and bounded  rationality in  a generalized  {\it Marriage
Problem}, comparing the benefits obtained by self-searching and by a matchmaker.
Finally  we  propose a
modified  matching   game  intended   to  mimic  the   way  consumers'
information makes  firms to enhance  the quality of  their products  in a
competitive market.
\end{abstract}
\end{frontmatter}

\section{Marriage model as a market metaphor}
The  Marriage Problem  \cite{GS} describes a  system  where two  classes of  $N$
players, that we  shall often call {\it men} and  {\it women}, can
be matched pairwise to their mutual benefit. 
In  different contexts players could be producers
and consumers, employers and job seekers or resources and activities in
general. In fact, while the metaphorical model of
marriages between men and women is suggestive, the importance of studying
matching models lies, for us, in their broad implications in economic and social contexts.
The fundamental premise is that there are many mutually beneficial relations
out there to be found, like the one linking consumers with specific
wishes to firms with suitable products. Mutually beneficial does
not imply equally beneficial: previous work shows that, 
whoever processes more information,
can in general reap more benefits \cite{happy}.

In this work we study further the matching model in some detail.
In particular, we want to compare the matchmaker mediated matches
(global optimum) with those obtained through the self-searching mode. 
The latter can be Nash equilibria \cite{nash} or {\it  satisficing}
matches, according to  the definition of  H. Simon
\cite{sim}. He  claims that people  face uncertainty about  the future
and costs in  acquiring information in the present. Sometimes information
is simply unavailable.
The best agents can do, with partial information and limited research 
capability, is setting an aspiration level which, if achieved, 
they will be happy enough with.

Even if agents were ``maximizers'' of some utility function,
partial information would still prevent potential optimal partners from finding
each other. Our analysis allows us to determine how good
are the approximations agents must be content with, in the absence
of complete information. This is particularly relevant in economic relations,
notably in the law of Supply and Demand. Neoclassical economics
generally assumes perfect information is available, leading to optimal solutions.
More recent researches, like the one conducted by Akerlof \cite{aker},
suggest that in extreme cases of asymmetric information, potentially
beneficial transactions may fail to materialize. 

However, when a transaction is still realized, even with
limited information, standard economics literature does not address
the problem of determining how far from the optimum it falls.
We try here, as it has previously been done in a different context \cite{Ebay},
to quantify the dependence of this distance from the amount of
available information. 
Our approach offers an open-ended research agenda, aiming
to give a micro detailed description of macroscopic matchings \cite{schell}.

\section{Average satisfaction in matching games}
Let us now formalize the model. We start with $2N$ players, 
$N$ men and $N$ women, who are to be matched.
Each player  is  been assigned  his/her  list of  preferred
partners. The lists  are drawn at random and  are independent from one
another. If man $m$ marries woman $w$ we attribute him an energy equal
to the ranking  of $w$ in $m$'s list.  Let us  define the matrices $f$
(for women) and $h$ (for  men), such that the element $f(w,m)$ denotes
the rank of man  $m$ in $w$'s list and $h(m,w)$ the  rank of woman $w$
in $m$'s list.   The average energy per person in  a given matching is
defined                     by                     
\be\label{epsilon1}
\epsilon=\frac{1}{N}\sum_{w=1}^{N}f(w,m_{\rm w})+\frac{1}{N}\sum_{m=1}^{N}h(m,w_{\rm m}),
\ee 
where $m_{\rm w}$ is the husband of  woman $w$ and $w_{\rm m}$ the wife of man
$m$. Low energy corresponds to high satisfaction and vice-versa.

There are  many ways to  match pairwise the  $2N$ players for  a given
instance  of  the preference  lists,  whether  or  not the  notion  of
stability  is taken  into account  \cite{Pit92}.  If  we  treat the
Marriage Problem  as an optimization  problem, then we can  define the
following relevant states:
\begin{itemize}
\item Ground State (state with the minimal total energy).
\item  Optimal  Stable State  (stable  state  with  the minimal  total
energy).
\end{itemize}
The first one is globally optimal, the second is a constrained optimal
state. Both of them are rather unlikely  to be found in real life, because of
incomplete information and  limited searching power \cite{dist}. In particular
men may not know all the women and vice-versa, or they may not dare to
divorce once they found an acceptable partner. 
Following the lines traced in ref. \cite{happy} we propose a new procedure, the {\it
First Choice}, that allows a  selfish agent to find a suitable partner
in a  limited time.  Here we shall mainly consider its  symmetrical version,
i.e.  we  make  no  difference  between  the  strategies  of  men  and
women. 

\begin{figure}
\epsfig{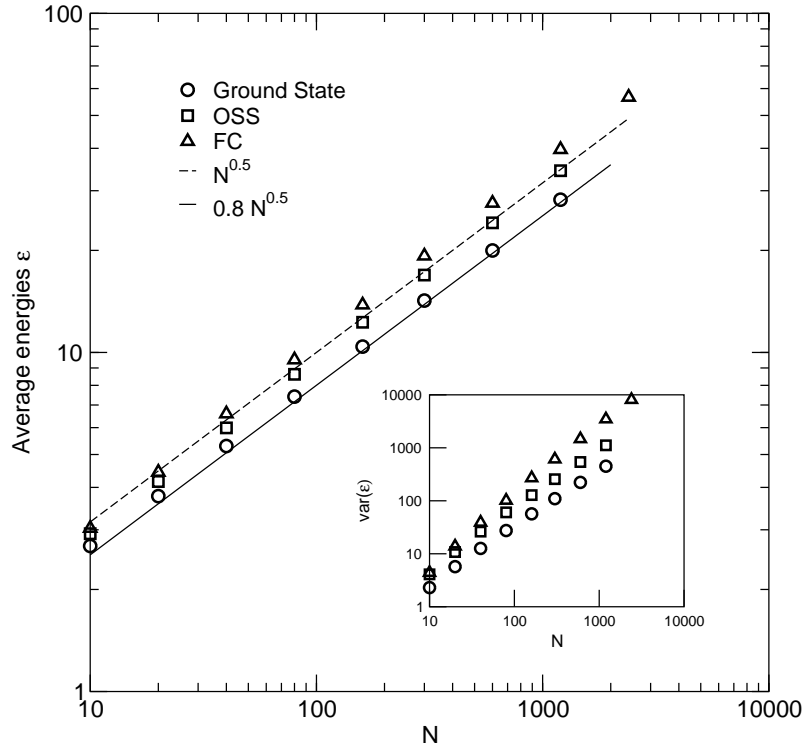} 
\caption{Average energy of agents $\epsilon$ as a function of $N$ in log-log scale. 
The dashed line is obtained from eq. (\ref{opt_avg_nrg}) and the solid line from eq. \req{gs_nrg}. 
Symbols correspond to numerical calculations of the Ground State (circles),
the OSS (squares) and the FC State (triangles), averaged over $1000$ repetitions.
In the inset graph we plotted their respective variances.} \label{fig1}
\end{figure}
\subsection{Ground State}
Often  referred to  as Assignment  Problem \cite{PS},  the  problem of
finding the matching of minimal energy in a bipartite graph has a long
history in Operations Research. It has been shown that, in an instance
of  size  $N$  of   the  marriage  problem,  the  Hungarian  algorithm
\cite{hun}  finds the  Ground  State in  $O(N^3)$  steps. The  average
energy of the corresponding  matching has been calculated analytically
in reference \cite{MZ}: 
\be\label{gs_nrg}
\epsilon^{GS}=0.81\sqrt{N}.
\ee
The above esteem (solid line) is compared with simulation data (circles)
in figure \ref{fig1}, with good agreement. 

\subsection{Optimal Stable State}
The Stable Marriage  Problem \cite{Knu76,GusIrv89} describes a complex
system where  individuals attempt to optimize  their own satisfaction,
subject  to mutually  conflicting constraints.   A stable  matching is
defined  by the property  that there  are no  two couples  $(m,w)$ and
$(m',w')$,  such  that  $h(m,w')<h(m,w)$  and  $f(w,m')<f(w,m)$.   The
algorithm of Gale and Shapley  \cite{GS} yields the 
stable  matching with minimal average men's energy ({\it male optimal}).  
Starting from the male optimal solution, all other stable matchings can 
be obtained by properly performing cyclic exchange processes called {\it rotations}.
If  one wanted  to  find the  stable matching  that
minimizes the total energy  $\epsilon$, one could thus enumerate all stable
matchings  and compare  their values.
Unfortunately it  has been
shown that the maximum and the average number of stable matchings,  for an
instance of size $N$ of the preference lists, grow, respectively, as $e^{N}$ 
\cite{IL} and as $N\log(N)$ \cite{GusIrv89,dierz}.  In order to find a faster algorithm 
reference \cite{K} suggests to construct a weighted directed graph of all rotations. The authors
show that the maximum-weight closed subset of the rotation poset gives
the OSS. The algorithm they  propose scales as $N^4$, i.e. the running
time of the Ford-Fulkerson algorithm  \cite{Sle} to find a minimum cut
in the graph.

Since in stable matchings the average male and female energies are found 
to obey the relation $\epsilon_{\rm m}\epsilon_f=N$ \cite{MZ}, in a sex-fair matching, such as 
the OSS, the average energy per person must be
\be\label{opt_avg_nrg}
\epsilon^{OSS}=\sqrt{N}.
\ee
In figure \ref{fig1} the above equation, plotted as a dashed line, 
is shown to fit well the simulation data we obtained for the OSS (squares) using 
the algorithm described in \cite{K}. 

\subsection{First Choice}
In the First Choice model (FC), at time step $n$ every player proposes
to the  $n^{th}$ of  his/her preference list.  Man $m$,  for instance,
will  propose to  woman $w$,  such that  $h(m,w)=n$. If  $w$  is still
unmarried, $f(w,m)\le n$  and $w$ receives no better  proposal at time
step $n$, then  she retains $m$'s one and  they get married. Marriages
cannot  be broken:  married  couples sit  aside  till the  end of  the
game. If there  are conflicting proposals the couples  with the lowest
energies are married first. When all the couples meeting this criteria
at time  step $n$ got  married, bachelors propose the  $(n+1)^{th}$ of
their preference lists, and so forth. The game proceeds till everybody
is married, reaching what we shall call FC state. 
Since energy corresponds to number of proposals, the FC state can be found in 
less than quadratic time.

The results of our extensive simulations, reported in fig. \ref{fig1},
show the gap between the three different ways of modeling agents'
behavior. The Ground State corresponds to the state of maximum
global satisfaction, but there is no reason why selfish players
should be able to find it by themselves, and not move from it once reached. 
Only an external institution, a matchmaker, 
could help people coordinate in such a profitable way. 
On the other hand, the OSS would not change, once reached by
rational players. Since there is no procedure to reach it, though, the same
rational players are much more likely to get trapped in another
stable state. The OSS is, then, the best matching perfect agents
could possibly obtain by self-searching.
Numerical simulations of the FC state, reported as triangles up in figure \ref{fig1}, 
show that the FC  matching gives  average energies  slightly above
those of the OSS, and the same holds for their
respective variances (inset graph). FC agents retain the best possibility they encounter
at a given time and do not need to process all the virtually infinite
information available. The same way consumers sometimes keep buying
a specific, satisficing, product, instead of spending their time and money
trying out all the other ones. 

\section{Individual energy distribution with complete information}
We turn now our attention to single individuals, by tracing the  individual energy
distribution for the three matching methods. That of the  OSS can be estimated
following the same reasoning as in reference \cite{dierz}, where 
the authors define the ensemble of  all stable matchings. In our case we 
just have to add the symmetry
constraint $\epsilon_{f}=\epsilon_{m}$. Using continuous variables and
rescaling the rankings such that
$x_{\rm m}\simeq h(m,w_{\rm m})/N$, $X=\sum_{i=1}^N x_i$ 
and $y_{\rm w}\simeq f(w,m_{\rm w})/N$, $Y=\sum_{j=1}^N y_j$, 
the probability of finding a symmetrical stable state reads:
\begin{eqnarray*}
P&=&\int_{0}^{1}  d^{N}x \int_{0}^{1}  d^{N}y  \prod_{i\neq j}  (1-x_i
y_j) \delta(\sum_i  x_i - \sum_j y_j) \\  &\simeq& 
\int_{0}^{1} d^{N}x e^{-X^2} \rho_{N}^{2}(X)
\end{eqnarray*}
where \cite{dierz}:
\begin{equation} \label{ro}
\rho_{N}(X)=\int_0^1   d^{N}x   \delta(X-\sum_{i=1}^{N}  x_i)   \simeq
\frac{X^{N-1}}{\Gamma(N)}(1-e^{-N/X})^N.
\end{equation}
The single agent energy distribution  can then be calculated by taking
the average of $\delta (x-x_N)$ in the above ensemble, which reads:
\begin{eqnarray}\label{robbacoatta}
<\delta(x-x_N)>&=&\frac{1}{P}\int_0^N e^{-X^2}\rho_{N-1}(X-x)\rho_{N}(X)\\
&\simeq& (N-1)\frac{(\sqrt{N}-x)^{N-2}}{N^{\frac{N-1}{2}}}
\rightarrow \sqrt{N} e^{-x\sqrt{N}},
\end{eqnarray} 
for $N \gg  x$. In fact
the integral in \req{robbacoatta} can be solved with the saddle point method
around $X=\sqrt{N}$, where the factor  $(1-e^{-N/X})^N$ of equation (\ref{ro})
is almost equal $1$.
Hence the distribution of  the average player's energy $\epsilon_{\rm i}=Nx$
equals 
\be\label{paolo}
p(\epsilon_{\rm i})=\frac{1}{\sqrt{N}}  e^{-\frac{\epsilon_{\rm i}}{\sqrt{N}}}.
\ee
The above becomes a simple exponential, if we rescale the
$\epsilon_{\rm i}$ variable as follows: 
\be\label{scen}  
\tilde{\epsilon_{\rm i}}=\frac{\epsilon_{\rm i}}{\sqrt{N}}.
\ee 
\begin{figure}
\epsfig{file=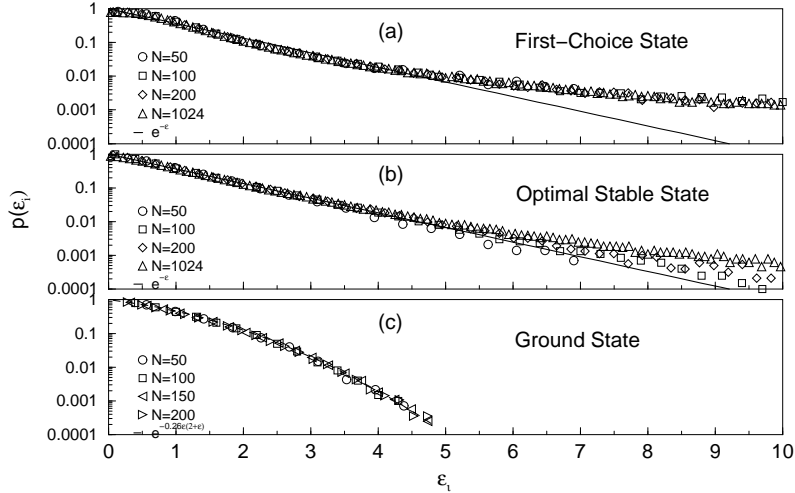, width=300pt} 
\caption{Probability distribution of average individual energies $\epsilon_{\rm i}$,
rescaled according to equation eq. \req{scen},
in the First Choice (a), 
Optimal Stable (b) and Ground State (c), plotted in linear-log scale. 
Symbols correspond to simulations obtained for different values of $N$, 
as reported in the legends. 
The solid lines represent the analytical solution for the OSS (equation (\ref{paolo})), 
the dashed line fits the data for the Ground State.} \label{fig2}
\end{figure}

Equation  \req{paolo}, plotted as a solid line,  
agrees with  the   numerical    simulations of the OSS  reported   in
fig.  \ref{fig2}.b, while it lies below the fat tail of 
the individual energy distribution of the FC State (fig. \ref{fig2}.a).
Such a discrepancy is due to the fact that FC marriages are unbreakable, so the
unlucky players married as last end up
with energies comparable  to that of a random  choice ($N/2$).
In real life social relations are not static, but local rearrangements,
such as divorces or lay-offs, take place on a much longer
time scale than that needed to find a partner.

The distribution of Ground State energies obeys perfectly
the  scaling of equation \req{scen} and  is   well  fitted  by   the  Gaussian-like
distribution $e^{-0.26\epsilon_{\rm i}(2+\epsilon_{\rm i})}$. This suggests that the globally
optimal solution tends to assign similar satisfactions to every agent, balancing
the loss of those who happen to be more ``beautiful'' \cite{capox} in a given instance of the
preference lists with the gain of the least attractive ones. Such a fair and
profitable matching, though, is prevented by the individual selfishness of
agents endowed with perfect information. In the following sections we will
show how things may change when individual information is incomplete.

\subsection{Evolutionary stability}
The model becomes  richer if we introduce a  threshold value $\delta$,
corresponding to the rank
above which it is more convenient for an agent to remain unmarried (or, similarly,
not to buy a given product).
In this situation one could wonder if it would be a better strategy to go
down quicker on one's preference list, thus lowering one's probability
of  remaining  single,  or slower,  in  order  to double one's chances 
of success with a better
partner. Let  us formalize  this idea as  follows: in a  population of
normal people  (who propose to one  partner at each time  step) we can
insert a percentage  $(1-\beta)$ of ``fast'' (who propose  to the next
two agents of  their preference list) or ``slow''  ones (who only make
one proposal every two time steps). The game can thence be carried out
as usual.  Some agents  will remain single  at an energy  cost $\delta
+1$.

\begin{figure}
\epsfig{file=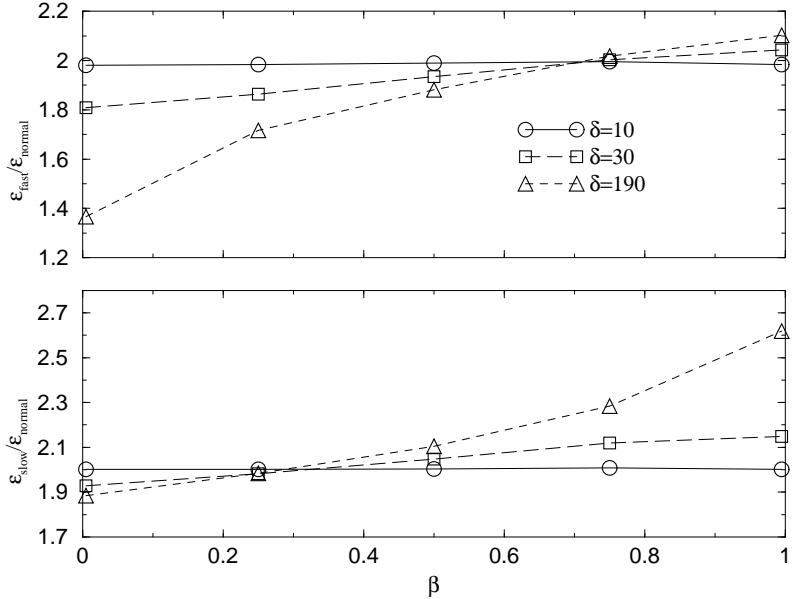, width=300pt} 
\caption{Ratio between ``fast'' (``slow'') people's energies and ``normal'' ones, 
as a function of the percentage of ``normal'' agents $\beta$. 
Symbols are simulation data for a $200$ agents system with different thresholds $\delta$. 
Lines are inserted to guide the eyes.} \label{slowfast}
\end{figure}
In   fig.  \ref{slowfast}   we  report   simulation  data from  this
model. There  we analyze  the ratios between  the average  energies of
slow      (fast)      and      normal      people      $\epsilon_{\rm slow
(fast)}/\epsilon_{\rm normal}$. Both ratios remain bigger than one for any
value of  the percentage  of normal people  $\beta$. This
proves that, given agents do not seek the optimal but the satisficing,
it is always convenient  to make only one  proposal at each time  step. 
In other words, we could make different possible strategies compete by Darwinian selection, 
giving each agent a number of offsprings proportional to her satisfaction 
at every generation. No matter the percentage of people holding
a given strategy in the first generation, the FC one would always spread and succeed.
In a wide sense  the First Choice model meets,  therefore, the requirements
for evolutionary stability \cite{may}.

\section{Partial information}
Complete information is rare to be encountered in economic and
social relations. Imperfect markets, where consumers do not even know the existence
of some products, are more frequent than ideal ones. 
The FC  model can be generalized  to the case  of limited information,
that is people only know a portion $\alpha$ of the world. Accordingly,
every  preference list  will contain  $N(1-\alpha)$ holes. If people behaved
according to the Gale-Shapley algorithm \cite{GS} the resulting satisfactions
would be those reported in reference \cite{dist}. 
In the FC case agents propose,  at time  step $n$,  to the  $n^{th}$  ranked known
counterpart player  of their lists.  Man $m$ could accept  woman $w$'s
proposal only  if $h(m,w)$ were  smaller or equal  to the rank  of the
$n^{th}$ known  woman of his list.  For small values  of $\alpha$ some
players  find no  partner and  they remain  single at  an  energy cost
$\Delta_c \geq N$.

It is evident that information matters: the more people one knows, the
more choice one has. It is  then easy to state that $\epsilon$ must be
a monotonically decreasing function of $\alpha$.
In  order to  quantify this  dependence,  we notice  that the  average
probability of  accepting a proposal is  of order $\epsilon  /N$. If a
player sent  out $n$ proposals  before finding her partner,  the total
probability $\epsilon\frac{n}{N}$ of  getting married must equal unity
(in the  absence of  singles). Starting from  this observation  we can
follow  the  argument  used  in  \cite{happy},  which  gives,  in  the
symmetric case,
\begin{equation} \label{zhang}
\epsilon=\sqrt{\frac{N}{\alpha(2-\alpha)}}.
\end{equation}
A similar, yet more precise esteem, can be found as follows.

\subsection{Analytical estimation}
\begin{figure}
\epsfig{file=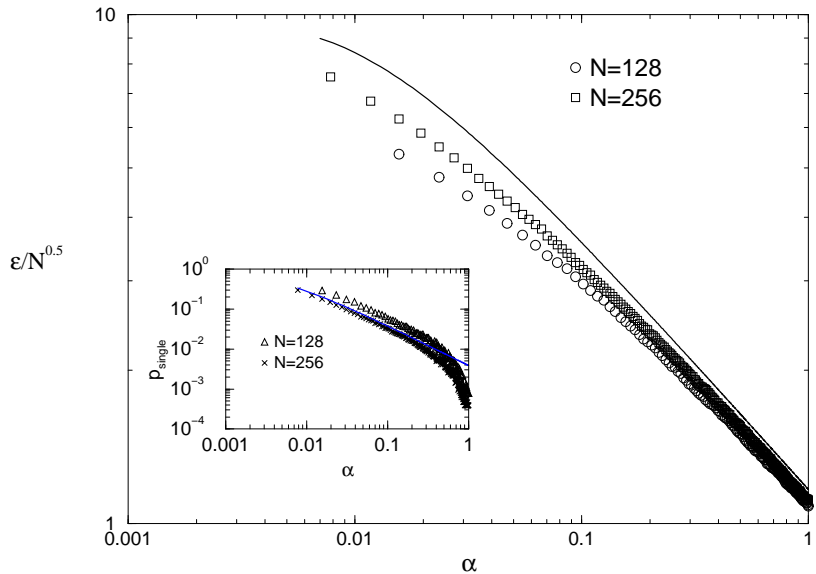, width=300pt} 
\caption{Simulations of the FC method at increasing $\alpha$ values, with $\Delta_c=N+1$. 
Circles ($N=128$) and squares ($N=256$) are the average energies of agents as a function of $\alpha$, in $\log-\log$ scale. The solid line plots equation \req{alfepsi} for $N=256$.
Inset: triangles up ($N=128$) and asterisks ($N=256$) are the percentage of singles, 
displayed in $\log-\log$ scale. The solid line plots $q_{\rm N\alpha}$ of equation 
\req{singles} for $N=256$.
} \label{fig0}
\end{figure}

The probability $p_n$ that an agent gets married exactly at time step $n$
can be written as
\be\label{pconn}
p_n\simeq 2\frac{n}{\alpha N} q^2_{n-1}, 
\ee
where $q_n$ is the probability of not being married and $n/\alpha N$ the average
probability of accepting a proposal, at step $n$. In eq. \req{pconn}, and in the following,
we are neglecting:
(i)The probability of having two agents proposing to each other at the same time.
(ii)The probability of having two agents proposing to the same partner at the same time.
The latter becomes important for $n_{max}=N\alpha\simeq 1$.
The number of singles diminishes with increasing $n$ according to the equation
\be\label{qmast}
q_n=q_{n-1}-p_n\simeq q_{n-1}\left( 1-2\frac{n}{\alpha N} q_{n-1} \right),
\ee
which can be solved for $\alpha\ll 1$, giving
\be\label{singles}
q_n=\left(1+\frac{n^2}{\alpha N} \right)^{-1}.
\ee
This means there are going to be singles even at the end of
the FC process, i.e. for $n=\alpha N$. We compared such $q_{\rm \alpha N}$
with numerical simulations in the inset of figure \ref{fig0},
finding a good agreement.
When $\alpha=1$ there are
no singles, and the exponential transition between the two regimes
takes place around $\alpha\simeq \log{N} /N$, which corresponds to the 
analogous cut-off found in marriages with threshold \cite{dierz}.

Among the players who get married at time step $n$, half of them
would be proposers and their energy would scale as $n/\alpha$; the
other half would receive a proposal uniformly distributed between
$1$ and $n/\alpha$. Therefore the average energy of these agents 
reads $\epsilon_n=\frac{3n}{4\alpha}$. 
Inserting equation \req{singles} into \req{pconn}, we can calculate
the average energy per agent as follows:
\be\label{alfepsi}
\epsilon=\sum_{k=1}^{N\alpha}p_k \epsilon_k
\simeq\frac{3\pi}{8}\frac{N\sqrt{N\alpha}}{1+N\alpha}.
\ee
The above equation is plotted against $\alpha$ 
in figure \ref{fig0}. Comparison with simulation data shows a better
agreement for large $\alpha$ values than for smaller ones.
This is probably due to the above mentioned fact 
that the neglected contribution of conflicting proposals 
becomes important when $n_{max}=N\alpha$ is small.

\subsection{Unhappy minority}
As  $\alpha$  increases,  the  average  energy per  person  drops  very
quickly. Now we ask ourselves how benefits are distributed
in the population. In particular we shall focus our attention
on the agents who do not benefit at all from an information
increase. This kind of individual based analysis is crucial
in economic systems, where the average data is not the end
of the story.

We can imagine a situation where agents are endowed with a 
knowledge $\alpha_{1}$, 
find a FC matching $M_1$ with average energy 
$\epsilon(\alpha_1)$, 
and then acquire an additional amount of
knowledge $\Delta\alpha$. 
This way players make new acquaintances and may find 
a better partner than they used to. The resulting new matching $M_2$ would 
correspond to a different point in the plot of fig. \ref{fig0}, i.e. $\epsilon(\alpha_2)$, 
where $\alpha_2=\alpha_1+\Delta\alpha$. The average energy decrease
$\epsilon(\alpha_2)<\epsilon(\alpha_1)$ is due to the percentage
$p_h(\alpha)$ of the population that improves its condition. 
On the other hand some players ($Np_u(\alpha)$) suffer the 
couples reshuffling from $M_1$ to $M_2$, ending up with a higher energy.

We  studied  the  percentage $p_h(\alpha)$ of ``happy'' people and that 
of unhappy ones $p_u(\alpha)$, at constant $\alpha$
increments  $\Delta\alpha$. 
\begin{figure}
\epsfig{file=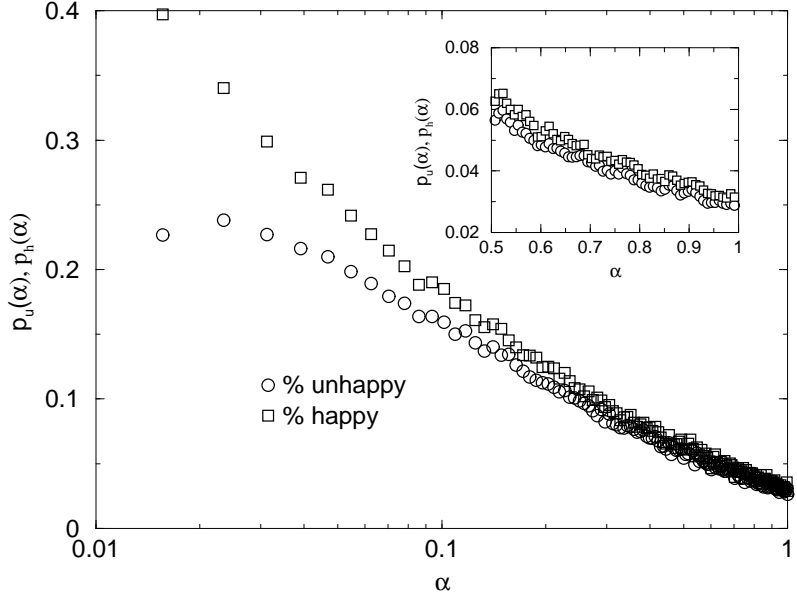, width=300pt} 
\caption{Simulations of the FC method at increasing $\alpha$ values, with $N=128$. 
Circles are the percentage of unhappy people $p_{\rm u}$, squares that of happy ones $p_{\rm h}$, 
at constant increments $\Delta\alpha=1/N$, plotted against $\alpha$ in semi-log scale. 
The inset highlights, in linear scale, the last half decade of the graph.} \label{unh}
\end{figure}
In figure  \ref{unh}  we reported  the
results for $\Delta\alpha=1/N$. Notice that, in this case, the derivative
$D_u(\alpha)=\lim_{\Delta\alpha\to 0}\frac{1}{N}\frac{p_u(\alpha)}{\Delta\alpha}$ 
if well estimated by $p_u(\alpha)$ itself, when $N$ goes to infinity.
After a very short transient phase where the number of singles increases
for  the joint  effect of  competition and  scarcity  of alternatives,
$p_u$  decreases logarithmically, and so does $p_h$. Notice that, even though
the two percentages come closer as $\alpha$ is increased,
$D_h(\alpha)$ stays always above $D_u(\alpha)$, as shown in the
inset of figure \ref{unh}. 

\subsection{Asymmetric information}
We have shown so far that information enhances the general satisfaction
and damages slightly only a small minority of the population. Let us now
give complete information to one side, say men ($\alpha_{\rm m}=1$), and vary 
the other group's one ($\alpha_{\rm w}\in(0,1]$). A similar situation
occurs in economics when, for instance, firms have a much
better knowledge of the market, that is their core business,
than consumers do. Potential buyers, on the other hand, may
increase considerably their amount of information by
gathering sellers' reputation data, which becomes
easily feasible in the Internet age \cite{Ebay}.

\begin{figure}
\epsfig{file=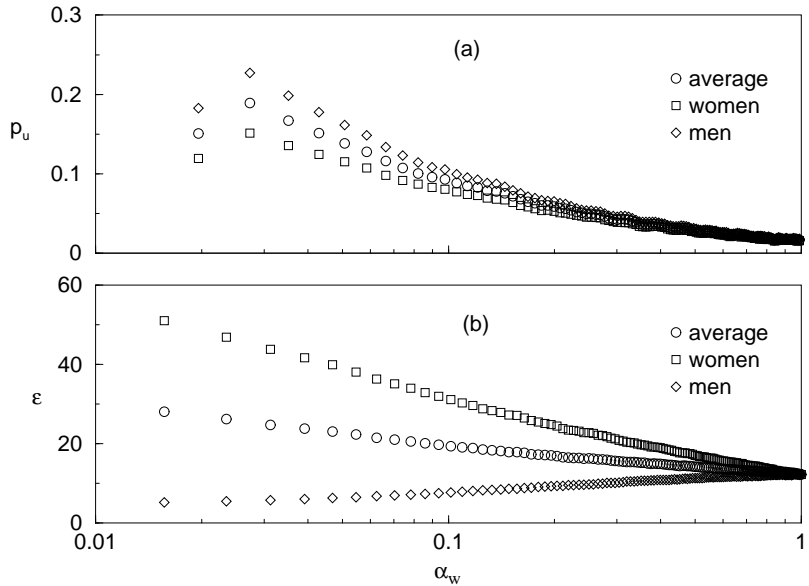, width=300pt} 
\caption{Simulations of the FC method at fixed $\alpha_{\rm m}=1$ and increasing $\alpha_{\rm w}$ values, with $N=128$. 
The lower graph displays the average energy of agents (men, women and both) as a function of $\alpha_{\rm w}$, in semi-log scale. 
In the upper graph the corresponding percentage of unhappy people (men, women and both), 
at constant increments $\Delta\alpha_{\rm w}=1/128$, 
is plotted against the same log-scaled $\alpha_{\rm w}$ values.} \label{asym}
\end{figure}

In figure \ref{asym}.a we plotted the probability of
being unhappy $p_u(\alpha_{\rm w})$, with the same definition as before, for
men, women and both. The behavior is very similar to the symmetric case.
As expected, there are more men than women suffering 
from the unilateral injection of information in the game.
Here men are justly paying the price of equal rights.

In figure \ref{asym}.b the corresponding energies of both sexes are 
plotted against $\alpha_{\rm w}$ in semi-log scale.
It appears clearly that women's energy (squares) decreases 
while men's energy (diamonds) increases, as $\alpha_{\rm w}$
grows. Then men's satisfaction deteriorates, even if their information
does not change during the process. Nevertheless the general average energy,
represented by open circles in figure \ref{asym}, diminishes. In other words
men's loss is more than compensated by women's improvement: even a unilateral
information increase is beneficial to the society as a whole.
This confirms the pie augmentation hypothesis made in \cite{happy}.

\section{Matchmaker}
When  information   is only partial  players
score very poorly (see fig. \ref{fig0}). In
some circumstances it could be convenient for them to let someone else
find them a partner, rather  than doing it by themselves. This becomes
possible  --and  even  likely--  in  a world  where  the  connectivity
distribution    of    human   social    contacts    is   very    broad
\cite{stanley}. More  than that, in  the Internet age very  few highly
connected nodes \cite{fal} could gather a huge amount of information.

Let  us  imagine  that people  in  a  community  agree to  give  their
preferences  lists to  a matchmaker  (MM)\cite{happy}. He  possess now
complete information and is capable of finding the Ground State of the
system.  Accordingly, he proposes  each player a partner: if everybody
accepted  the MM's  proposal, then  the community  would  minimize the
total energy. Nevertheless we  assume players are selfish and careless
of the  general welfare. Thus they  keep looking for  a better partner
among those players whom  they know, employing their maximum searching
effort.  If a man  $m$ and  a woman  $w$ prefer  to marry  one another
rather than accepting the MM's offer, then they refuse it, step out of
the community and get married.  Such a decision forces in the meantime
two more players ($m$'s and $w$'s partners in the Ground State) to look
for  a new  partner.  For  each  value of  $\alpha$ there  would be  a
percentage $p_i(\alpha)$ of such ``independent'' people.

\begin{figure}
\epsfig{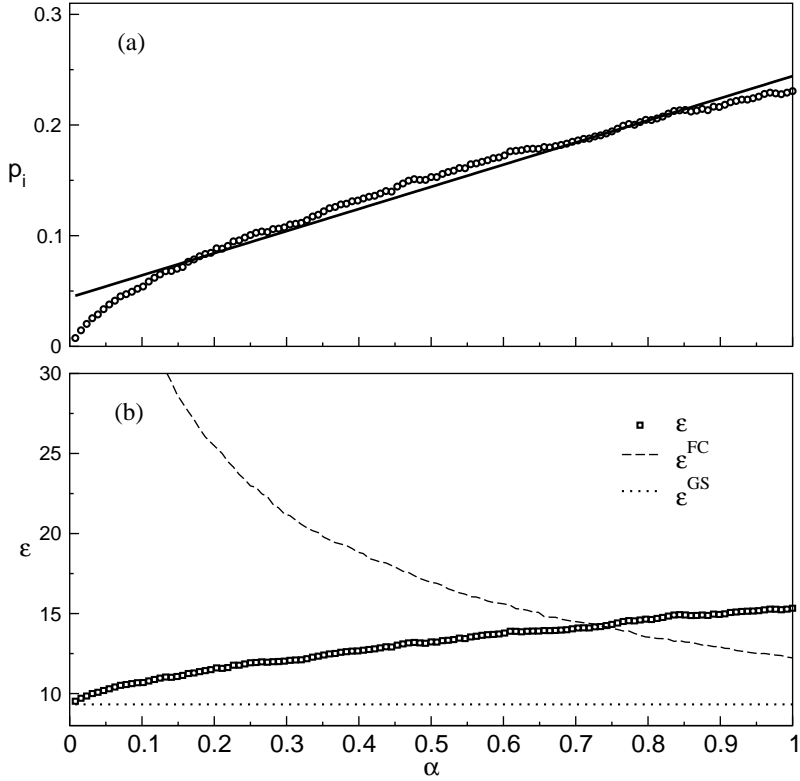} 
\caption{Simulation data of a Marriage game with $N=128$ players for each gender. Averages are taken over $10^4$ realizations for each value of $\alpha$, whose x-axis linear scale is the same for all graphs. 
(a)Circles represent the percentage $p_{\rm i}$ of independent people. 
The solid line plots equation \req{mat}. 
(b)Squares represent the overall average energy per person $\epsilon$. 
The dotted line is the value of the Ground State energy, corresponding to $p_{\rm i}=0$, the dashed line that of the FC energy.} \label{match}
\end{figure}

A rough  esteem of $p_i$ can be  obtained by calculating
the probability that an individual FC energy is lower than
that of the Ground State. To do that, we shall assume that
the distribution of individual energies in the FC state $p^{FC}(x,\alpha)$ equals
that of the OSS (eq. \req{paolo}) for $\alpha=1$, and that 
$p^{FC}(x,1)dx=p^{FC}(x/\sqrt{\alpha},\alpha)dx/\sqrt{\alpha}$. 
These approximations are only
reasonable when $\alpha \gg \alpha_c$. If we take equation \req{gs_nrg}
for the individual energy distribution in the Ground State,
we obtain:
\be\label{mat}
p_i(\alpha)\propto \left[ \sqrt{\alpha} \int_0^{\sqrt{N}} dx e^{-x} \int_0^{x} dy e^{-0.3 y(2+y)} \right]^2
\simeq 0.2 \alpha,
\ee
for $N$ approaching to infinity.
In  figure  \ref{match}.a  equation  \req{mat}  is
compared with simulations, and  the  agreement seems good.

In  figure  \ref{match}.b  the   overall  average  energy  per  person
$\epsilon$ is plotted against $\alpha$. Its lower and upper bounds should
respectively be the  Ground State energy  (displayed as dotted  line) and
the FC one  (displayed as dashed line). 
It is clearly not the case for the latter, because half of the independent
couples are rematched randomly, having been forced to divorce by their
mates. The crossing point (around $\alpha=0.72$) corresponds, in the graph 
\ref{match}.a, to a critical
percentage of independent people. Above that value (about $2\% $) the MM
can no more be useful, unless he employed sophisticated mechanisms. For
instance he could find the matching of minimal energy among the ``loyal'' people; 
but this could become an
endless process, since it would generate a new cascade of desertions,
and so forth. 

For sure, at least below that critical value,
$\epsilon$  is  a monotonically  increasing function  of
$\alpha$.  This  means  that  people with  partial  information  (small
$\alpha$) can  score better than  those with infinite  searching power
($\alpha=1$),  given  that  they   renounce  to  their  personal  {\it
privacy}. This effect would be amplified by the fact that some of them
would find the MM's offer  satisficing, and not search any further for
a  different  partner.   In  other  words  what  we   are  showing  in
fig. \ref{match}.b  is the worst  case of anti-social behavior  at any
given average  information level. In  real life very few  people would
employ  all  their searching  effort  looking  for  a substitute  good
(partners might  be an exception)  if they were satisfied  enough with
the one  they already  have \cite{sim}.  This  way most of  the people
will remain  in the  MM's community, thus  saving searching  costs and
being better  off on average.  Such savings can  be thought of  as the
MM's revenue. The above argument  becomes more effective if we imagine
that the  game be  repeated numerous times,  in which  case reputation
matters \cite{Ebay}.  Then people  would care more about their average
utility and might be pleased by a long lasting confidence relationship
with the MM.

Of course  we should consider the  possibility of a  corrupted MM, who
would favor some players at the expense of the collectivity.  However,
the matchmaker institution would survive only if the average energy of
those who accept his proposals is lower than that of those who found a
partner  on their own.   Moreover, when  the information  that players
release  to  the  MM is  not  used  properly,  it  would start  to  be
convenient for  them to  protect their privacy.   In such  a situation
it's no more profitable for the  MM to be corrupted and an equilibrium
is likely  to arise.   By construction, at  the equilibrium  point the
average satisfaction  is bigger than it  would be without  the MM, and
more   evenly  distributed   (the   variance  of   the  Ground   State
distribution,   plotted  in  the inset of fig. \ref{fig1},  is   the  smallest).
Furthermore, we  should consider  competition among different  MMs, in
which case  a cheating one would  be exposed to the  danger of loosing
clients and lower his earnings.

\section{Market implications} 
A real  market could be thought  of as a polygamic  marriage game with
unequal number of players. In order to point out the role of information,
we can try to carry on this metaphor letting rational agents with
partial information play the game. 

Let us  consider a  market with $N_c$  consumers and  $n$ enterprises,
such that $n\ll N_c$, where every enterprise produces the same type of
good.  Enterprises  have  a  completely  degenerate  preference  list,
i.e.  they  make no  discrimination  among  different consumers.  Each
enterprise  $i$ is  endowed with  a budget  equal to  $1$,  a fraction
$(1-q_i)$  of which  will  be  invested in  the  marketing process,  a
fraction  $q_i$  in  the  quality  of the  product.  Here  ``quality''
includes    the    research   expenditures    and    the   costs    of
production. Consumers  act as in  the FC model.  They get to  know the
different brands  through commercials, plus an  effort $e_k\in (0,1)$,
which represents the amount of personal search. $e_k=1$ corresponds to disposing of and processing
infinite   information,  $e_k=0$   to  no   information   effort.  The
probability  $\alpha_k(i)$ that  consumer  $k$ knows  the product  $i$
should be a convex function  of $(1-q_i)$; for the sake of simplicity,
let us assume it linear:
\begin{equation}\label{pubbli}
\alpha_k(i)=\min \{1,1-q_i+e_k\}.
\end{equation}
On the  other hand, for consumer  $k$ the product $i$  has an expected
utility
\begin{equation}\label{exut}
u_k(i)=q_i+\eta_k (1-q_i),
\end{equation}
where  $\eta_k$ is  a random  variable, uniformly  distributed between
zero and one.  This way consumers have a perception  of the quality of
the goods($q_i$), but commercials ($1-q_i$) are aimed to confuse them,
so  that  the  products look  nicer  than  they  really are.  In  fact
information is not only limited in  real life, but it is also imperfect
\cite{happy,Ebay}.  The   factor  $\eta_k$  is  intended   to  account  for
commercials more or less effective,  for different tastes and needs of
consumers. Higher  expected utility  corresponds to better  ranking in
one's   preference  list.   This  approach   resembles   the  ``beauty
correlated''   marriage   problem,   studied   in   a   recent   paper
\cite{capox}.  All  products  have   the  same  price,  but  different
qualities.

\begin{figure}
\epsfig{file=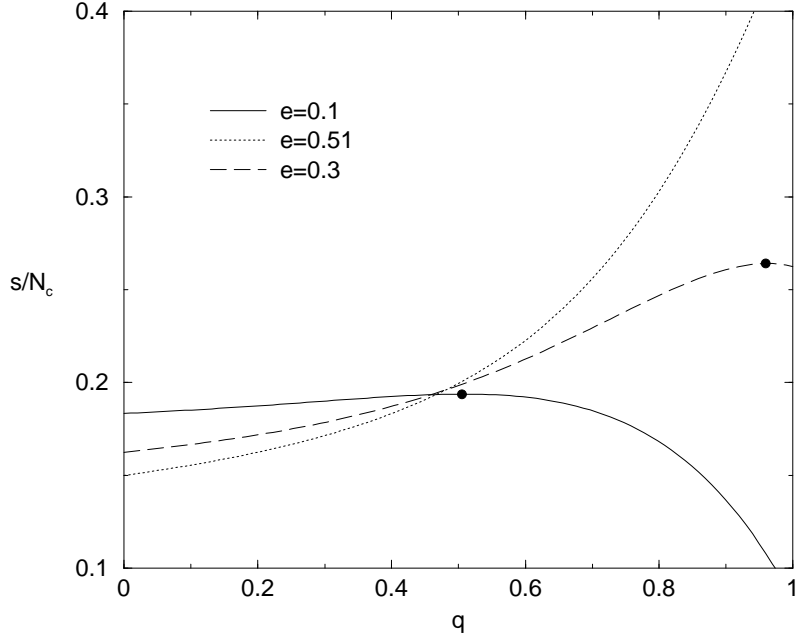, width=300pt} 
\caption{Probability of selling an item as a function of the quality investment $q$. 
The plots represent eq. (\ref{ef}), with $n=10$, for three different 
values of the average consumers' effort $e$. Filled points are the maxima $\hat{q}$ 
of the curves for $e=0.1$ and $e=0.3$. For $e=0.51$ $s$ is an increasing function of $q$, 
therefore $\hat{q}=1$.} \label{effort}
\end{figure}

At each instance of the game  consumers buy one unit of good. In order
to maximize their expected utility,  they choose the best ranked among
those they know of.  As a first, simple example, let us  set up a game
with  the  variables  $q_i$  randomly  distributed  between  zero  and
one.  Each consumer  knows  an average  of $r(e)=n(1+2e)/2$  different
products if the average effort $e$ is smaller than $0.5$, $n$ products
otherwise.  Consumers' average utility  is a  monotonically increasing
function of  $e$ and $n$.  Now we introduce  a new enterprise  and ask
ourselves  the $q$ value  $\hat{q}$ that  maximizes its  selling. With
some approximation we can show  that the probability of being the best
ranked,  among the  $r(e)$  that  each consumer  knows,  has the  form
$p_{best}(q,r)\simeq      \frac{(1-q^{r-1})}{(1-q)r}$.     Multiplying
$p_{best}$ and  $\alpha_k(i)$ (eq.  \ref{pubbli}) we find  the average
selling of the new enterprise:
\begin{equation}\label{ef}
s(q,e)                              =                              N_c
(1-q+e)\frac{(1-q^{n\frac{1+2e}{2}-1})}{(1-q)n\frac{1+2e}{2}}
\end{equation}
For $e<0.5$  we can assume $r\simeq  n/2$. With this  position we find
that  the   maximum  of   the  above  equation   lies  on   the  curve
$$e(\hat{q})=\frac{(n-2)(1-\hat{q})^2
\hat{q}^{n/2}}{2\hat{q}^2+\hat{q}^{n/2}[\hat{q}(n-\hat{q})+2-n]},$$
that is  a monotonically increasing function of  $\hat{q}$. This means
that,  as  expected,  the   optimal  fraction  of  quality  investment
$\hat{q}$   increases   as    the   consumers'   effort   grows   (see
fig. \ref{effort}). 
In  other words, the bigger portion  of the market
consumers  know, the  more enterprises  are  likely to  invest in  the
quality of the product. Moreover,  the amount of effort needed to make
it convenient to  the new enterprise to invest all  its capital in the
quality,   $e(\hat{q}=1)=\frac{4}{n-4}$,   diminishes  sensibly   with
increasing $n$. Similar arguments hold for $e>0.5$.

At a given value of  the consumers' effort $e$, some enterprises would
adjust  their $q$ investment  to exploit  the edge  left by  the other
ones.  But  how do  consumers  decide  their  searching effort  level?
Increased  computing  power   may  allow  for  increasing  information
capability  with  no additional  cost.  Still, some information is  costly
\cite{stig}.  One way to  take care  of it  is subtracting  the search
costs to the  expected utility. In this more  realistic case consumers
may decide how much they want to invest in searching a good, given the
average quality  of the products.  In the scenario we  pictured above,
they  would   probably  search  until  they   are  satisfied,  without
necessarily maximizing their utility function. In fact, what is a core
business for an  enterprise is a marginal one  for most consumers. For
this reason, even  if some enterprises had a  perfect knowledge of the
market and could find their $\hat{q}$  at any time, it would always be
convenient for them to chose a lower level of quality investment.

\section{Acknowledgments}
We would like to thank A. Capocci and Joseph Wakeling for useful comments.
This work was supported by the Swiss National Fund, Grant No. 20-61470.00.

\end{document}